\documentclass[pra,twocolumn,aps,amsmath,showkeys,showpacs,floatfix,nofootinbib]{revtex4-1}

\usepackage{graphicx}
\usepackage{dcolumn}
\usepackage{amsfonts}
\usepackage{bm}
\usepackage{color}

\newcommand{\ket}[1]{\left|#1\right\rangle} 
\newcommand{\bra}[1]{\left\langle#1\right|} 
\newcommand{\braket}[2]{\left< #1 \vphantom{#2} \right|
 \left. #2 \vphantom{#1} \right>} 
\newcommand{\op}[1]{\ensuremath{\Hat{\mathrm{#1}}}}

\def\vec#1{{\boldsymbol{#1}}}  

\begin{document}


\title{Quasihole dynamics as a detection tool for quantum Hall phases}

\author{ T. Gra\ss$^1$, B. Juli\'a-D\'{\i}az$^{1}$, and M. Lewenstein$^{1,2}$}

\affiliation{$^1$ICFO-Institut de Ci\`encies Fot\`oniques, Parc Mediterrani 
de la Tecnologia, 08860 Barcelona, Spain}
\affiliation{$^2$ICREA-Instituci\'o Catalana de Recerca i Estudis Avan\c cats, 
08010 Barcelona, Spain}

\begin{abstract}

Existing techniques for synthesizing gauge fields are 
able to bring a two-dimensional cloud of harmonically 
trapped bosonic atoms into a regime where the occupied 
single-particle states are restricted to the lowest 
Landau level (LLL). Repulsive short-range interactions 
drive various transitions from fully condensed into 
strongly correlated states. In these different phases 
we study the response of the system to quasihole 
excitations induced by a laser beam. We find that 
in the Laughlin state the quasihole performs a coherent 
constant rotation around the center, ensuring conservation 
of angular momentum. This is distinct to any other 
regime with higher density, where the quasihole is 
found to decay. At a characteristic time, the decay 
process is reversed, and revivals of the quasihole 
can be observed in the density. Measuring the period 
and position of the revival can be used as a spectroscopic 
tool to identify the strongly correlated phases 
in systems with a finite number of atoms.
\end{abstract}

\pacs{67.85.De,73.43.-f}
\keywords{Quantum Hall states. Collapse and revival. Artificial gauge fields. Ultracold bosons.}
\maketitle

\section{Introduction}

Strong correlations and anyonic excitations are the 
intriguing properties of quantum states in two-dimensional 
systems exposed to strong magnetic fields. They show up  in the context of
fractional quantum Hall effect of electrons~\cite{laughlin,arovas}. In recent 
years, the advances in techniques for cooling and 
controlling atoms have raised the hope that these 
interesting states might also be artificially generated 
in systems of ultracold atoms~\cite{cooper-wilkin-gunn}. 
This would allow to experimentally confirm the fundamental 
theoretical concept of fractional quantum statistics~\cite{par2}, 
and open the door for topological 
quantum computation~\cite{nayak}. 

The key requirement for realizing such states is a 
strong external gauge field, which due to the 
electroneutrality of the atoms has to be synthesized. 
Artificial gauge fields which are strong enough to 
bring the system into a regime, where only the 
lowest Landau level (LLL) is occupied, have already 
been generated by rotating a gas of $^{87}$Rb~\cite{schweikhard}. 
The occurrence of strongly correlated states in 
the LLL regime then crucially depends on the ratio 
between trapping energy, favoring condensation in 
states with small angular momentum, and the strength 
of repulsive interactions, which tends to spread the 
atoms over a wide range of angular-momentum states. 
In a system of bosons interacting via a two-body 
contact potential, this competition is known to 
restrict the Laughlin state~\cite{laughlin} to a 
narrow region of parameters of extremely weak 
effective trapping, and thus close to the instability 
at the centrifugal limit~\cite{cooper-aip, smith00, bruno-njp}. 
This drawback has so far hindered the experimental 
realization of the Laughlin state. It has led to the 
proposal of using laser-induced geometric phases to 
mimic magnetic fields (cf.~\cite{dalibard}). Such a 
method, experimentally proven in Ref.~\cite{lin}, 
allows for a precise tuning of the gauge field strength 
as required for reaching the Laughlin state. An experimental 
route to produce the Laughlin state could start with preparing 
the system in a condensate at zero angular momentum, 
$L=0$. Then, stepwise transitions into states with higher
angular momentum can be induced by adiabatically increasing the 
gauge field strength~\cite{brunoPRA,bruno-njp}, until reaching 
the bosonic Laughlin state, characterized by $L=N(N-1)$ 
(in units of $\hbar$) with $N$ the particle number.

An important question is then how to detect this state. 
Its zero compressibility or its constant bulk density 
are characterizing features, but do not uniquely distinguish 
the Laughlin state from other quantum liquid states. 
Moreover, in systems of only few particles these attributes 
may become quite unsharp, while experimental progress 
in realizing Laughlin states of few particles has 
been reported~\cite{gemelke}, and even small systems 
have been predicted to support bulk properties like 
fractional excitations~\cite{bruno-njp}. Thus, looking 
for distinctive features, experimentally accessible even 
in small clouds, seems to be expedient. 

In this paper, we discuss a scheme for testing many-body 
quantum states in the LLL by piercing a quasihole into 
them. Experimentally, this can be achieved by focusing 
a laser beam onto the atomic cloud. After switching off 
this laser, the subsequent dynamics of the quasihole 
can be observed in the density of the system. We show that 
it yields relevant information about the underlying 
state. The defining property of the Laughlin state, being 
the densest state with zero interaction energy in a 
two-body contact potential, is found to be reflected in 
a decoherence-free dynamics of the quasihole. This is 
in clear contrast to the time evolution of a quasihole 
pierced into a state with $L < N(N-1)$.  In this case, 
an interaction-induced dephasing delocalizes the excitation, 
visible in the density as a decay of the quasihole. We 
explicitly consider a quasihole in the $L=0$ condensate, 
and in a Laughlin-type quasiparticle state. For these 
states we show, that the decay process is reversed 
at a characteristic time, leading to a revival of the 
quasihole.

This dynamics is reminiscent of the collapse and revival 
of a coherent light field which resonantly interacts 
with a two-level atom. This effect has been studied 
theoretically in the framework of the Jaynes-Cummings 
model since the early 1980s~\cite{eberly80,averbukh92}, 
and has experimentally been observed in systems of Rydberg
atoms~\cite{walther,stroud,brune96}, or trapped ions~\cite{wineland}. 
With the realization of a Bose-Einstein condensate 
(BEC) in 1995, also interacting many-body systems 
have become candidates for studying such collapse-and-revival 
effects: In Ref.~\cite{maciek96} it has been argued 
that quantum fluctuations cause a phase diffusion which 
leads to a collapse of the macroscopic wave function. 
As a consequence of the discrete nature of the spectrum, 
periodic revivals of the macroscopic wave function have 
been predicted in Refs.~\cite{wright96,imamoglu97}. It 
has been proposed to produce macroscopic entangled
states by time-evolving a condensed state~\cite{sorensen01,brunooberthaler}. 
An interesting scenario has been discussed in 
Refs.~\cite{castin97,wright97}, studying collapse and 
revival of the relative phase between two spatially separate 
BECs. Measuring phase correlations between many
BECs which are distributed on an optical lattice has 
allowed for observing the collapse and revival of 
matter waves~\cite{Collapse1}. Recently, the observation 
of quantum state revivals has been proven to provide 
relevant information about the nature of multi-body 
interactions in a Bose condensed atomic cloud~\cite{will10}.

Also the collapse and revival which we discuss in this 
paper allows to extract useful information: The effect 
itself not only clearly distinguishes the Laughlin 
regime from denser ones, but also measuring the revival 
times and positions of the quasiholes allows to 
determine the kinetic and interaction contribution 
to the energy of the system.

Our paper is organized in the following way: After 
introducing the system in Sec.~\ref{system}, we study the coherent quasihole
dynamics in the Laughlin state in Sec.~\ref{laugdyn}. This is in
contrast to the collapse-and-revival dynamics in denser regimes described in
Sec.~\ref{CR}. 
In Sec.~\ref{conc}, we draw our
conclusions.

\section{The system \label{system}}

We consider a two-dimensional system of bosonic atoms 
with mass $M$, described by the effective Hamiltonian 
${\cal H}= \sum_{i=1}^N H_i + {\cal V}$, where the 
single-particle contribution reads
\begin{align}
\label{Hi}
  H_i = 
\frac{(\vec{p}_i+\vec{A}_i)^2}{2M}  +  \frac{M}{2} \omega_{\rm eff}^2(x^2+y^2).
\end{align}
Here, $\vec{A}_i$ denotes the artificial gauge potential 
acting on the $i$th particle. It shall describe a gauge 
field of strength $B$ perpendicular to the system. We 
choose the symmetric gauge, $\vec{A}_i = \frac{B}{2} (y_i,-x_i,0)$. 
Different proposals for synthesizing this gauge potential 
are reviewed in Refs.~\cite{cooper-aip,dalibard}. The 
trapping potential is affected by the generation of 
the gauge potentials, but it is possible to make the 
effective trap axial-symmetric with trapping frequency 
$\omega_{\rm eff}$. It is useful to introduce a quantity 
$\omega_{\perp} \equiv \sqrt{\omega_{\rm eff}^2 + \frac{B^2}{4M^2}}$, 
which in the case of a rotation-induced gauge field 
equals the applied trapping frequency. From now on it 
will be used to fix units of energy, $\hbar \omega_{\perp}$, 
and units of length $\lambda_{\perp}=\sqrt{\hbar/(M\omega_{\perp})}$. 

The first term in Eq.~(\ref{Hi}) is seen to give rise 
to Landau levels. Using the dimensionless parameter 
$\eta \equiv B/(2M\omega_{\perp}) \leq 1$, we can 
express the Landau level gap as $\Delta_{\rm LL}=2\eta$. 
The degeneracy of states in each level is split by 
the second term in $H_i$. In the LLL, the eigenenergies 
are given by $E_{\ell}=(1-\eta)\ell+\epsilon_0$, 
corresponding to the Fock-Darwin (FD) states 
$\phi^{\rm FD}_{\ell}(z) \propto z^{\ell} \exp(-|z|^2/2)$, 
with $z=x+iy$, and $\ell$ the single-particle angular 
momentum. The term $\epsilon_0$ describes an $\ell$-independent 
zero-point energy. The interaction ${\cal V}$ is assumed 
to be repulsive $s$-wave scattering, described by
\begin{align}
{\cal V} = \frac{\hbar^2 g}{M} \sum_{i<j} \delta^{(2)}(z_i-z_j),
\end{align}
where $g$ parametrizes the interaction strength such 
that $gN/4\pi$ equals the mean-field interaction energy 
per particle. To avoid populating higher Landau levels, 
we need to fulfill $gN/8\pi \ll \eta<1$. Within this 
constraint, we can freely tune either $g$ via Feshbach 
resonances, or $\eta$ by modifying the gauge field 
strength. This allows to drive the many-body ground 
state from a condensate 
\begin{equation}
\Psi_0(z_1,\dots,z_N) \propto \prod_{i=1}^N \phi^{\rm FD}_0(z_i)
\end{equation}
for $\eta \ll 1$, to the Laughlin state 
\begin{equation}
\Psi_{\rm L}(z_1,\dots,z_N) \propto \prod_{i<j} (z_i-z_j)^2 \exp(-\sum_i
|z_i|^2/2)
\end{equation}
for $\eta \rightarrow 1$~\cite{brunoPRA}.

After cooling the system into the ground state of 
$\cal H$, we use laser beams to pierce quasiholes 
into the state. In Refs.~\cite{par2,bruno-njp}, it has 
been confirmed by exact diagonalization that a laser 
potential focused at position $\xi$, 
$V_I \propto I \sum_i \delta^{(2)}(z_i-\xi)$, is able to 
produce quasihole excitations in the Laughlin state. 
The laser intensity $I$ needs to be strong enough 
to close the gap protecting the Laughlin state. 
Defining a quasihole operator 
$\op{O}_{\rm qh}(\xi) \equiv {\cal N} \prod_i (z_i-\xi)$ 
which pushes away all particles from position $\xi$ 
and where ${\cal N}$ re-normalizes the state, the 
quasihole state corresponding to a state $\Psi(z_1, \dots, z_N)$ 
can generally be defined as 
$\Psi_{{\rm qh}}(\xi;z_1,\dots,z_N) = \op{O}_{\rm qh}(\xi) \Psi(z_1, \dots, z_N)$. 
This state has up to $N$ units of 
angular momentum more than the original state.

If the system is large enough, our procedure can be 
repeated in order to create several distinct quasiholes. 
In the following we investigate the scenario where, after generating the
quasiholes, the laser beams are abruptly switched off. In general, a dynamical
evolution is expected, since $\Psi_{\rm qh}$ is not an
eigenstate of $\cal H$. This is directly clear for $\xi \neq 0$, where the
operator $\op{O}_{\rm qh}(\xi)$ breaks the cylindrical symmetry of the
Hamiltonian $\cal H$. We restrict our study to quasiholes characterized by 
a vanishing wave function at $\xi$, thus we do not consider the possibility of
quasiholes of a different form, as for instance the half-flux excitations in the
Moore-Read state~\cite{moore-read}.

\section{Coherent quasihole dynamics in the Laughlin state \label{laugdyn}}

First, we consider the Laughlin state with one quasihole 
at position $\xi$:
\begin{equation}
\label{qhsum}
\Psi_{{\rm L,qh}}(\xi) \equiv \op{O}_{\rm qh}(\xi) \Psi_{\rm L} 
= {\cal N} \sum_{k=0}^N \xi^{N-k} f_k(z_1,\dots,z_N) \Psi_{\rm L},
\end{equation}
where the $f_k$ are totally symmetric polynomials of $k$th 
order in the coordinates $z_1, \dots, z_N$, with the property 
that each of the coordinates appears at most to linear order. 
Since ${\cal V}\Psi_{\rm L}=0$, we also have ${\cal V} f_k \Psi_{\rm L} = 0$. 
Furthermore, all $f_k \Psi_{\rm L}$ are homogeneous polynomials 
in the $z_i$ times the overall Gaussian, and thus are 
eigenstates of the single-particle part $\sum_i H_i$ with 
eigenvalue $E_k = (1-\eta)L_k +N \epsilon_0=(1-\eta)[k+N(N-1)]+N \epsilon_0$. 
Defining $\gamma \equiv N(N-1)(1-\eta) + N \epsilon_0$ and 
$\epsilon \equiv (1-\eta)$, we can write the time evolution 
of the quasihole state:
\begin{equation}
\Psi_{\rm L,qh}(\xi,t) 
\equiv \mathrm{e}^{\frac{i}{\hbar}{\cal H}t} \Psi_{\rm L,qh}(\xi) 
= {\cal N} \mathrm{e}^{\frac{i}{\hbar} \gamma t} 
\sum_{k=0}^N \mathrm{e}^{\frac{i}{\hbar} \epsilon k} \xi^{N-k}  f_k \Psi_{\rm L}.
\end{equation}
The exponential in the sum can be absorbed by making the 
$\xi$'s time-dependent: 
$\tilde \xi(t) \equiv \xi \mathrm{e}^{-\frac{i}{\hbar} \epsilon t}$. 
With this we obtain:
\begin{align}
\label{qhdyn}
\Psi_{\rm L,qh}(\xi,t) & 
= {\cal N}  \mathrm{e}^{\frac{i}{\hbar}(\gamma + N \epsilon) t} 
\sum_{k=0}^N \tilde \xi(t)^{N-k} f_k \Psi_{\rm L} \nonumber \\
&= \mathrm{e}^{\frac{i}{\hbar}(\gamma+N\epsilon) t} \Psi_{\rm L,qh}(\tilde \xi(t)).
\end{align}
The overall phase evolution, $\mathrm{e}^{i(\gamma + N \epsilon)t}$, 
is just the dynamics of a state with a quasihole in the center, 
$\xi=0$, which is an eigenstate of $\cal H$. As Eq.~(\ref{qhdyn}) 
shows, for a symmetry-breaking quasihole off the center, the 
conservation of angular momentum is ensured by a constant 
coherent rotation of the quasihole around origin. Any change 
in $|\xi|$ is forbidden by conservation of angular momentum. 
The angular velocity of the quasihole is given by 
$\omega_{\perp} (1-\eta)$. Remarkably, no decoherence between 
different terms in the sum of Eq.~(\ref{qhsum}) occurs 
during the evolution. This phenomenon is a consequence of 
the state's zero interaction energy in a contact potential. 
Note that this behavior is in contrast to conventional 
fractional quantum Hall systems with long-range interactions 
where a tunneling of the quasihole to the edge is 
expected~\cite{hu}. 

The above calculation can easily be repeated for more 
than one quasihole. It can generally be shown that the 
time dependence of the corresponding wave function can 
be absorbed into the positions of the quasiholes 
$\xi_i \rightarrow \xi_i  \mathrm{e}^{-\frac{i}{\hbar} \epsilon t}$, 
and an overall phase factor. Therefore we just have to 
note that the normalization factor of the wave functions, 
which carries the information about the anyonic 
statistics of the quasiholes, depends only on absolute 
values $|\xi_i|$ and $|\xi_i-\xi_j|$, thus the 
substitution $\xi_i \rightarrow \xi_i \mathrm{e}^{-\frac{i}{\hbar} \epsilon t}$ 
poses no problem there.

\section{Collapse and revival of the quasihole \label{CR}}
In contrast to the coherent dynamics of quasiholes in the Laughlin state,
 in this section we will encounter collapse-and-revival processes of
quasiholes pierced into phases of less angular momentum, $L<N(N-1)$.

\begin{figure*}[t]
\centering
\includegraphics[width=0.28\textwidth,angle=-90]{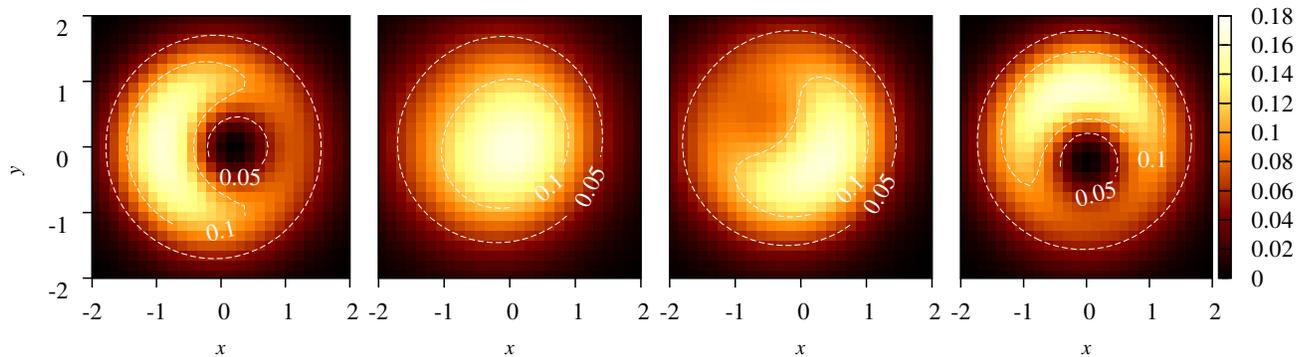}
\caption{ Contour plot of the density (divided by $N$): 
(a) of a $L=0$ condensate with one quasihole at $\xi=0.2$ at $t=0$
($N$-invariant), 
(b) at time $t=T/2$ for $N=8$ [see definition of $T$ in Eq.~(\ref{T}], 
(c) at $t=T/2$ for $N=20$, (d) at $t=T$ ($N$-invariant).}
\label{Fig1}
\end{figure*}
\subsection{Quasihole in the $L=0$ condensate \label{L0dyn}}
We start analyzing the dynamical behavior of a 
quasihole in a $L=0$ condensate, described by the wave 
function $\Psi_{0,\rm qh}(\xi) = \op{O}_{\rm qh}(\xi) \Psi_0$. 
As before in the Laughlin case, we can decompose this 
expression into a sum over homogeneous polynomials, 
$\Psi_{0,\rm qh}(\xi) \propto \sum_{k=0}^N f_k \xi^{N-k} \Psi_0$, 
where every term is an eigenstate of the single-particle 
part of $\cal H$, with corresponding eigenvalue 
$k \epsilon + \gamma'$. Now, $\gamma'=N \epsilon_0$ is the zero-point 
energy. We should bear in mind that both $\epsilon$ and $\epsilon_0$ depend on
$\eta$, so their numerical values might be different to the Laughlin case.

Again, we can absorb the single-particle contribution 
being linear in $k$ into the time evolution of the 
quasihole, so for the non-interacting system, 
$g=0$, we would have 
\begin{equation}
\Psi_{0,\rm qh}(\xi,t) = 
\exp\left[\frac{i}{\hbar} (N\epsilon + \gamma') t\right] 
\Psi_{0,\rm qh}(\tilde \xi(t))\,,
\end{equation} 
in full analogy with Eq.~(\ref{qhdyn}). Interactions, 
however, change the situation: The terms 
$\varphi_k \equiv f_k \Psi_0$ are in general not 
eigenstates of the interaction $\cal V$. To describe 
the time evolution of this system, we thus have to 
decompose the $\varphi_k$'s into an eigenbasis of 
$\cal V$. Since $\cal V$ conserves angular momentum, 
we can restrict ourselves, for every $\varphi_k$, to 
the subspace with $L=k$, for which we obtain the 
eigenbasis via exact diagonalization. We denote this 
basis by $\chi^{(k)}_{\alpha}$ and write:
\begin{equation}
 \varphi_k = \sum_{\alpha} c^{(k)}_{\alpha} \chi^{(k)}_{\alpha}.
\end{equation}
The coefficients $c^{(k)}_{\alpha}$ can easily be obtained: 
The exact diagonalization yields the $\chi^{(k)}_{\alpha}$ 
in the Fock basis of occupation number states 
$\ket{n_{\ell=0}, n_{\ell=1}, \dots}$. In this basis, the state 
$\varphi_k$ is represented by the vector $\ket{N-k,k,0,\dots}$, 
from which it differs only by a normalization factor 
$N_k = [ (N-k+1)! \ (k+1)! ]^{-1/2}$. We thus have 
$c^{(k)}_{\alpha} = N_k \braket{\chi^{(k)}_{\alpha}}{N-k,k,0,\dots}.$ 
The $\chi^{(k)}_{\alpha}$, being homogeneous 
polynomials of $k$th degree, are eigenstates of the kinetic 
term with the eigenvalue $k\epsilon + \gamma'$. We now may write
\begin{equation}
\label{Phidecomp}
 \Psi_{0,\rm qh}(\xi,t) = 
\mathrm{e}^{\frac{i}{\hbar} (N\epsilon +\gamma') t} 
\sum_{k=0}^N \tilde \xi(t)^{N-k} \sum_{\alpha} 
c^{(k)}_{\alpha} \chi^{(k)}_{\alpha} \mathrm{e}^{i \epsilon_{\alpha}^{(k)} t}.
\end{equation}
Here, $\epsilon_{\alpha}^{(k)}$ is the eigenvalue of 
$\cal V$ corresponding to the eigenvector $\chi^{(k)}_{\alpha}$. The presence 
of this term causes, in general, a dephasing of the 
different contributions to $\Psi_{0,\rm qh}(\xi)$. Thus, 
while the single-particle contribution just rotates the 
quasihole at fixed radial position $|\xi|$, the interaction 
makes the quasihole fade out, as shown in the density 
plots of Fig.~\ref{Fig1}. Also, slight deformations of the 
cloud as a whole become apparent during the time evolution, 
in clear contrast to the Laughlin case. This interaction-driven 
dynamics also happens for a symmetry-conserving quasihole in 
the center.

\begin{figure}[t]
\centering
\includegraphics[width=0.48\textwidth]{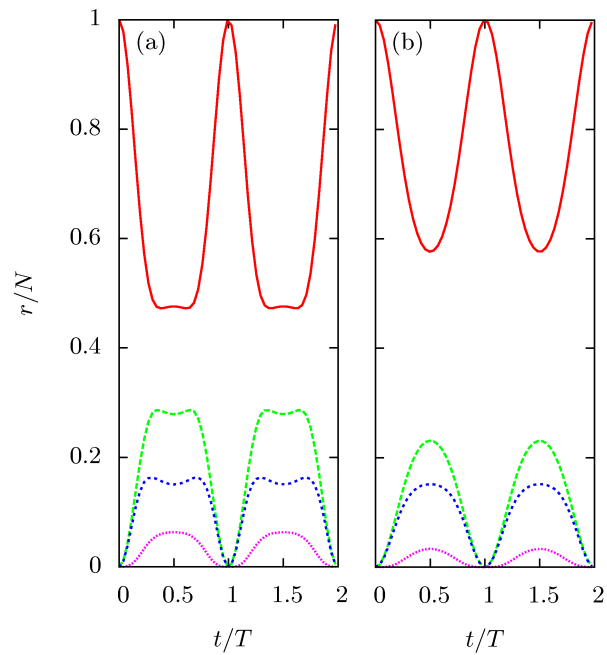}
\caption{Eigenvalues $r$ of the one-body density matrix $\rho$ 
as a function of time. Initially, one hole was pierced 
at $\xi=0.2$ into the $L=0$ condensate.  (a) $N=8$. 
(b) $N=20$.}
\label{Fig2}
\end{figure}

To quantify the dephasing we consider the eigenvalues $r$ 
of the one-body density matrix 
$\rho_{ij} = \bra{\Psi_{0,\rm qh}(\xi,t)} \op{c}_i^{\dagger} \op{c}_j \ket{\Psi_{0,\rm qh}(\xi,t)}$, 
that is, the occupation of the different eigenmodes. Here, 
$\op{c}_i$ ($\op{c}_i^{\dagger}$) is the annihilation (creation) 
operator of a particle in FD state $\phi_i^{\rm FD}$. As 
shown in Fig.~\ref{Fig2}, initially all particles occupy 
the same single-particle state, $(z-\xi) \exp(-|z|^2/2)$, 
and all eigenvalues of $\rho$ are zero except for 
one which is equal to the number of particles $N$. If 
the hole is placed in the center, the different 
instantaneous eigenmodes of $\rho$ are the Fock-Darwin 
functions, as a matter of fact that the initial state 
has definite angular momentum, and the Hamiltonian 
conserves angular momentum. For all particle numbers 
we have studied, the most occupied mode is, at any 
time, $\phi_{1}^{\rm FD}$, while the second most occupied 
is the Gaussian, i.e. $\phi_{0}^{\rm FD}$. As can be 
inferred from Fig.~\ref{Fig2}, and as we have verified 
explicitly by calculating the pair-correlation 
function~\cite{grassPhD}, the time 
evolution brings the initially fully uncorrelated system 
into a correlated state, where detection of one particle 
at position $z$ influences the outcome of another measurement 
at position $z'$. The effect is most pronounced in small 
systems with a quasihole in the center. In this case, we find 
for $N=12$ at $t=T/2$ a correlated state with almost equally 
populated modes, $n_0^{\rm FD} = 4.7$ and $n_{1}^{\rm FD}=4.3$.

As seen in Figs.~\ref{Fig1} and~ \ref{Fig2}, the collapse of 
the quasihole is followed by a perfect revival. We thus have 
oscillations between fully condensed quasihole states and 
correlated states. In the absence of perturbations, these 
oscillations of period $T$ will continue forever. Note 
that, in general, the position of the quasihole at the 
$n$th revival, $\xi^{(n)}$, changes from period to period due 
to the kinetic contribution. In the following, we will 
derive expressions for $T$ and $\xi^{(n)}$, which will 
allow to deduce information about different system's 
parameter by measuring these quantities.

Therefore, we have to analyze the spectra 
$\epsilon_{\alpha}^{(k)}$, which define the periods 
$\tilde T_{\alpha}^{(k)} = 2\pi \hbar / \epsilon_{\alpha}^{(k)}$ 
of the phase oscillations of each contribution 
$\chi_{\alpha}^{(k)}$. Now assume that there are some 
energy units $u$ and $u'$, which allow to write
\begin{eqnarray}
\label{epsspec}
\epsilon_{\alpha}^{(k)} = k u' + n_{\alpha}^{(k)} u,
\end{eqnarray}
with $n_{\alpha}^{(k)} \in \mathbb{N}$. The part linear 
in $k$ does not depend on $\alpha$, and thus can be 
absorbed into the kinetic energy contribution. It 
exclusively affects the revival position by defining 
the time dependent parameter $\tilde \xi(t)$, now 
rotating with an angular velocity $(\epsilon+u')/\hbar$. 
The phase decoherence between different $\alpha$ 
states is controlled by $u$, and the corresponding 
periods read $T_{\alpha}^{(k)} = 2\pi \hbar (n_{\alpha}^{(k)}u).$ 
We see that $T=2\pi\hbar/u$ is a multiple of all 
$T_{\alpha}^{(k)}$, so at time $T$, all contributions will 
have the original phase relations.

To determine $u$ and $u'$, we numerically analyze 
the spectra $\epsilon_{\alpha}^{(k)}$. We find that the 
gap above the $L=N$ subspace provides us, for 
any $N$, with an energy unit $u \approx 0.040 gN.$ 
All states in the spectrum are found to be given as 
integer multiples of $u$ plus the ground-state energy. 
Also in subspaces $L \leq N$, the same unit $u$ can 
be used to quantize most of the energies. Strikingly, 
the eigenstates to energies which cannot be 
constructed according to Eq.~(\ref{epsspec}) have 
zero overlap with $\varphi_k$. Thus, they do not 
contribute in Eq.~(\ref{Phidecomp}).

The second energy unit of Eq.~(\ref{epsspec}) turns 
out to have exactly the same value, $u'=u$. As 
exact solutions are known for the ground state energies 
in subspaces with $L\leq N$~\cite{bertsch99,jackson00,smith00}, 
we can write down an analytic expression $u'=gN/8\pi$. We 
thus obtain for the revival period [in units of $\omega_{\perp}^{-1}$]:
\begin{equation}
\label{T}
 T = \frac{(4 \pi)^2}{gN}.
\end{equation}
From this formula, we directly see that choosing 
$gN={\rm constant}$ makes the oscillation periods 
independent from the size of the system. This choice 
is convenient as it also guarantees a finite interaction 
energy per particle in the thermodynamic limit. 
Fixing $g$ rather than $gN$, the periods would 
decrease in larger systems. By Eq.~(\ref{T}), a 
measurement of the revival period $T$ directly 
yields information about $gN$. Measuring then 
the polar angle $\phi^{(n)}$ of the revival position 
$\xi^{(n)} = \tilde \xi(nT)$ will allow to 
extract $\eta$. We find
\begin{align}
 \phi^{(n)} = \frac{4n (1- \eta)}{gN}+n.
\end{align}

The effect of the system size can be seen by 
comparing $N=8$ and $N=20$ at $t=T/2$ in Figs~\ref{Fig1} 
and~\ref{Fig2}: The larger the system, the more it tends to 
maintain its initial properties. In Refs.~\cite{bertsch99,jackson00,smith00,ueda06}, 
it has been shown that the ground state wave functions for 
$L \leq N$ are closely related to the functions $f_k$, 
which are the wave functions where the total angular 
momentum $L=k \leq N$ is most equally distributed 
amongst $N$ particles, 
$f_k = \sum_{1 \leq i_1 < \cdots < i_k \leq N} z_{i_1} \cdots z_{i_k}$. 
From these functions, we obtain the polynomial part of 
the ground state wave function of $L=k$ by replacing 
the coordinates $z_i$ by the relative coordinates 
$\tilde z_i \equiv z_i -Z$, where $Z$ is the 
center-of-mass coordinate. Note that $Z$ is not 
just a number, but an operator with $\langle Z \rangle =0$. 
As center-of-mass fluctuations decrease with 
increasing particle number of the system, for 
large-sized systems, $Z$ becomes pinned to 
the center, and the states $f_k \Psi_0$ become 
eigenstates of the Hamiltonian with eigenvalues 
$E_k \propto k$. Just like in the Laughlin case, we will 
then no more observe the collapse and revival of the hole. 
The rotational movement around the origin 
will survive the thermodynamic limit if the hole is 
initially placed outside the center. Therefore, 
the dynamics of a single hole does not qualitatively 
distinguish the condensed phase from the Laughlin 
phase in the thermodynamic limit. There is, however, 
a quantitative difference, as the period of rotation 
will be shorter than the period of Laughlin quasiholes 
due to the energy $u'$ from Eq.~(\ref{epsspec}) which 
has to be absorbed in the definition of $\tilde \xi(t)$. 

\subsection{Piercing two quasiholes in the condensate\label{twoholes}}

\begin{figure}
\centering
\includegraphics[width=0.48\textwidth]{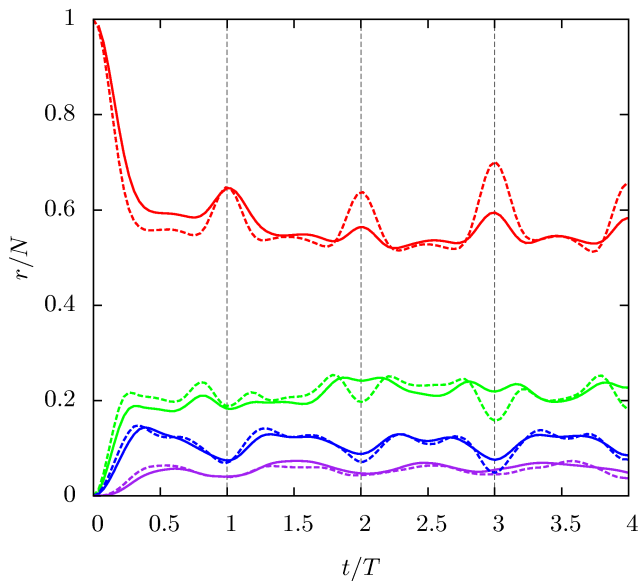}
\caption{Occupation numbers $r/N$ of eigenmodes as a 
function of time in a condensate of $N=8$ (dashed lines) and 
$N=12$ (solid lines) particles. Initially one hole has been 
placed in the center and second hole at $|\xi|=1$. No 
full revivals are observed, and the peaks at $t=nT$ 
are washed out in larger system. \label{Fig3}}
\end{figure}

The situation becomes quite different if we pierce 
a second hole into the condensate. For simplicity, 
we choose to introduce one of them in the center. Our 
wave function is then a linear combination of states 
$\ket{0,N-k,k,0,\cdots}$ with $0 \leq k \leq N$, and 
we are in the angular momentum regime with $N \leq L \leq 2N$. 
Here, even the GS energies $E_L$ do not behave linearly 
with $L$~\cite{cooperwilkin}. Also, many more states 
are involved when expressing the states 
$\ket{0,N-k,k,0,\cdots}$ in terms of eigenstates of 
the interaction, which now live in a significantly 
increased Hilbert space. The consequence of this is that, 
after a quick dephasing, the holes will never exhibit 
a full revival, see Fig.~\ref{Fig3}. Contrarily, as 
the comparison of the data for $N=8$ and $N=12$ in 
Fig.~\ref{Fig3} suggests, the peaks at the same period 
as given by Eq.~(\ref{T}) are expected to fade away 
quickly in the thermodynamic limit.

An $L=0$ condensate with two quasiholes is similar 
to an $L=N$ condensate with one quasihole, and 
in the thermodynamic limit the $L=N$ condensate 
with one vortex in the center is the ground state 
of the system at this angular momentum. Fig.~\ref{Fig3} 
therefore suggests that in regimes $L\gtrsim N$, 
also a single quasihole will dephase. This seems 
to be reasonable, as spontaneous symmetry breaking 
has been predicted for states with $L\gtrsim N$~\cite{cooper-aip,dag,dag2}, 
and ground states in several $L$ subspaces become 
quasi-degenerate. That means that energy differences within each
quasi-degenerate manifold  are very small, while large 
energy jumps occur between different manifolds. Therefore, a 
dephasing of different contributions and thus a 
collapse of the quasihole must be expected. Furthermore, 
the smallness of energy contributions in the quasi-degenerate 
manifolds make revival times unobservably long. We do 
not further investigate this situation of $L\gtrsim N$, 
as the symmetry-breaking leads to the formation of 
vortex lattices~\cite{cooper-aip}, as observed in 
experiments~\cite{schweikhard}. Characteristics of 
the lattice should allow for a clear identification of these phases.

\subsection{Dynamics of a quasihole in the Laughlin quasiparticle state
\label{LQPdyn}}

\begin{figure}
\centering
\includegraphics[width=0.48\textwidth]{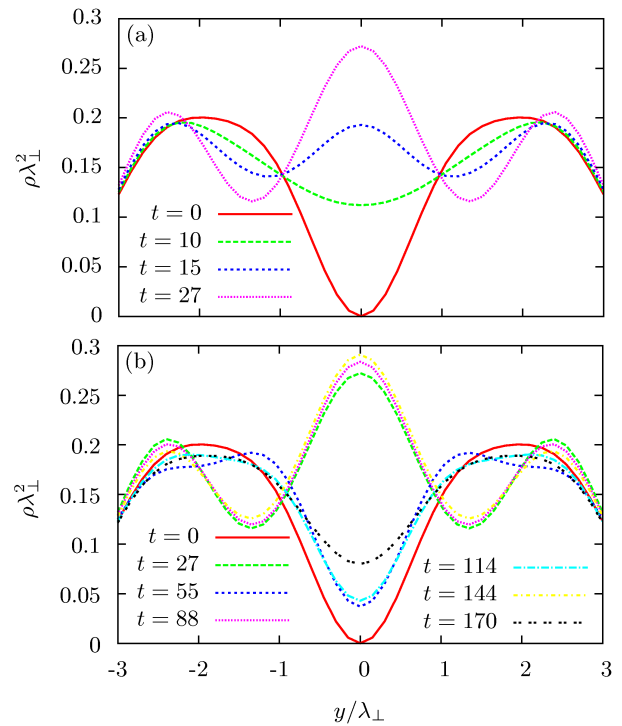}
\caption{Density profile ($x=0$) for a system of $N=6$ particles 
initially prepared in the state $\op{O}_{{\rm qh}}(0) \Psi_{\rm L,qp}$ 
at different times $t$ (in units of $\omega_{\perp}^{-1}$) with $g=1$. 
In (a) the evolution from a quasihole in the center to a density 
peak is shown. In (b) we plot the density for all times $t<200$, 
at which it takes an instantaneous extremum in the center.
\label{Fig4}}
\end{figure}

Upon increasing the gauge field strength, the vortex lattice has 
been predicted to melt when reaching filling factors 
$\nu \lesssim 6$~\cite{cooper-wilkin-gunn}. Then, a variety of 
strongly correlated quantum liquid phases are candidates for 
the ground state. Finally, for $\nu=1/2$ or $L=N(N-1)$, the 
Laughlin state becomes the ground state of the system. It 
is certainly in this regime where observable properties 
to distinguish between the phases become most relevant. 
Let us therefore study the dynamics of a quasihole pierced 
in the last incompressible phase which has been predicted 
to occur before reaching the Laughlin state. It is 
characterized by $L=N(N-2)$ and by a wave function which 
differs from the Laughlin wave function only locally at 
the origin, having the form of a Laughlin quasiparticle 
excitation, $\Psi_{\rm L,qp} = \exp(-\sum_i |z_i|^2/2) \partial_{z_1} \dots 
\partial_{z_N} \prod_{i<j} (z_i-z_j)$~\cite{bruno-njp}. 

For simplicity, we will pierce the quasihole in the origin, 
which makes the resulting state an eigenstate of the 
single-particle part of ${\cal H}$, and all dynamics will 
exclusively be driven by ${\cal V}$. To obtain the state 
$\op{O}_{{\rm qh,0}} \Psi_{\rm L,qp}$ in the Fock basis, we 
numerically diagonalize the Hamiltonian 
${\cal H}' = {\cal V} + V_I$ in the subspace $L=N(N-1)$. 
We then decompose this state in the corresponding eigenbasis 
of ${\cal V}$, also obtained by exact diagonalization. 
Several eigenstates of ${\cal V}$ will contribute, but 
the largest contribution comes from the Laughlin state 
with an overlap of 0.709 (0.717) for $N=7$ ($N=6$). 
Expressed in the eigenbasis of $\cal V$, we can easily 
perform the time evolution of $\op{O}_{{\rm qh}}(0) \Psi_{\rm L,qp}$.

The dynamics is clearly visible in the density shown in 
Fig.~\ref{Fig4}a: In the course of time, the hole fades 
out, as the center of the cloud gains a finite density. 
At some point, even a density maximum is developed at the 
origin, surrounded by a circular density valley. As the 
valley spreads out, the maximum becomes clearly peaked. The 
process is then reversed, and a hole at the center re-appears. 
Such oscillations between a density maximum and a density 
minimum in the center can be observed repeatedly. The scenario, 
however, differs from the collapse-and-revival process in 
the condensate: First, re-appearing holes are not equivalent 
to the original hole, as their density at the center remains 
finite, and their core size has decreased, see 
Fig.~\ref{Fig4}b. Second, the ``revival'' periods are not 
sharp. In Fig.~\ref{Fig4}b, we have chosen precisely those 
times at which the process is reversed. For $N=6$, the 
reversal after the first re-appearance of the quasihole is 
found at $t = 55$, while the second revival takes place at $t=114$.

\begin{figure}
\centering
\includegraphics[width=0.48\textwidth]{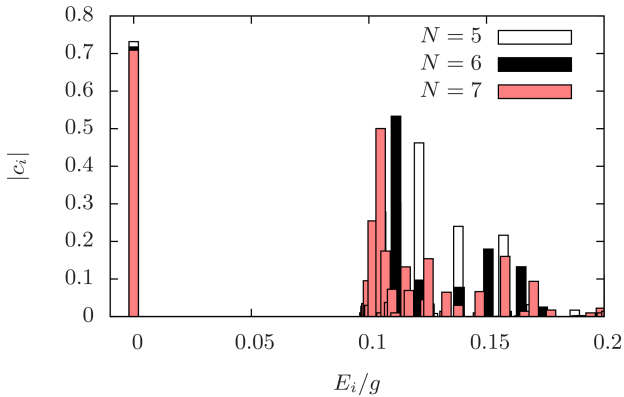}
\caption{Overlap $c_i$ between $\op{O}_{{\rm qh}}(0) \Psi_{\rm L,qp}$ and 
eigenstates of $\cal V$ with energy $E_i$ for different $N$. \label{Fig5}}
\end{figure}

To understand this behavior, we have to analyze the 
spectrum of $\cal V$ at $L=N(N-2)+N=N(N-1)$. It can be 
divided into a quasi-continuous excitation band and 
the Laughlin state. A gap $\Delta$ separates these two 
contributions. For $N \gtrsim 6$, the gap approaches the 
constant value of $\Delta \approx 0.1g$, if we choose $g$ (rather than 
$gN$) to be constant~\cite{Regnault:2003, bruno-njp}. 
Compared to this value, the energy differences between 
states within the excited band are typically very small. 
This property of the spectrum can be seen in Fig.~\ref{Fig5}, 
where we have plotted the overlap of the eigenstates 
with the initial state, $c_i \equiv \langle \op{O}_{{\rm qh}}(0) \Psi_{\rm
L,qp} | \chi_i \rangle$, versus $E_i/g$. Here, $\ket{\chi_i}$ denotes 
the eigenstates of $\cal V$ in the $L=N(N-1)$ subspace, 
and $E_i$ is the corresponding eigenenergy.

Due to this structure, the relative dephasing of different 
contributions to $\op{O}_{{\rm qh}}(0) \Psi_{\rm L,qp}$ from 
the excited band is slow compared to the dephasing of 
these contributions with respect to the contribution 
from the Laughlin state. Following this reasoning, 
$2\pi/\Delta \approx 63$ for $g=1$ sets the rough 
time scale for a ``quasi''-revival, at which the 
Laughlin state is again ``in phase'' with the low-energy 
contributions from the excited band. This slightly 
differs from the number we find by analyzing the 
density (cf. Fig.~\ref{Fig4}b), $t=55$ for $N=6$. But 
we note that the most important contribution from 
the excited band is a state with energy $E=0.112g$ 
(cf. Fig~\ref{Fig5}). Thus, it is in phase with the 
Laughlin after a time $t=56$. It has an overlap with 
$\op{O}_{{\rm qh}}(0) \Psi_{\rm L,qp}$ of 0.533. A 
superposition of this state and the Laughlin state 
is able to reproduce the quasihole state with a 
fidelity of 78\%. Other important states, with 
overlaps 0.279 and 0.110, are found at $E=0.105g$ 
and $E=0.102g$. At $t=56$ they are still nearly 
in phase with the $E=0.112g$ state. But also states 
with $E=0.164g$ and $E=0.182g$ contribute significantly 
with overlaps 0.180 and  0.133. These states will 
be clearly out of phase, making the revival imperfect. 
Subsequent revivals will more and more suffer from 
the slow dephasing within the manifold of excited 
states. This explains the small irregularity in 
the revival periods and the loss of the quasihole 
character in the density profiles  (see Fig.~\ref{Fig4}b).

Posing the question whether the described dynamics will 
survive in the thermodynamic, we first note that due to 
the similarity between the Laughlin state and 
$\op{O}_{{\rm qh}}(0) \Psi_{\rm L,qp}$, we can always 
expect contributions from both the Laughlin state 
and the excited band. This assessment seems to 
agree with Fig.~\ref{Fig5}, where the overlap with 
the Laughlin state is found to be almost constant 
while varying particle number, $5 \leq N \leq 7$.
As the Laughlin gap is known to be constant for large 
$N$, also the revival time has to be. The imperfection 
of the revival, characterized by a finite density at 
the center, continuously improves, as we increase 
the system size from $N=4$ to $N=7$. This seems 
reasonable, since energy differences in the excited 
band decrease with larger $N$, slowing down the 
dephasing in the excited band. In Fig.~\ref{Fig5} 
this reflects in the decreased spreading of relevant 
states when increasing particle number. However, whether 
this might lead to a perfect collapse-and-revival 
process in the thermodynamic limit, is not clear from our 
calculations.

\section{Conclusions \label{conc}}

We have shown that observing the dynamics of a quasihole 
might serve to classify different ground states in the 
LLL regime. Especially, the absence of decoherence is 
a characteristic feature of the Laughlin regime, 
$L \geq N(N-1)$, due to its zero interaction energy. It
is in clear contrast to the collapse of the quasihole 
observed in finite systems with $L<N(N-1)$. For 
simplicity, we have considered an idealized system 
with a cylindrically-symmetric Hamiltonian. We note 
however that even deformed Laughlin states, as 
expected in laser-induced gauge fields~\cite{brunoPRA,bruno-njp}, 
should be characterized by a decoherence-free 
quasihole dynamics due to their vanishing interaction 
energy. In the condensed regime, the collapse of 
the quasihole is followed by a perfect revival, 
and the system oscillates between a condensed and 
a correlated state. System parameters like $gN$ 
and $\eta$, specifying the interaction and the 
single-particle energy, can be obtained by measuring 
period and positions of this revival. A collapse-and-revival 
is also found for a quasihole in the Laughlin-quasiparticle 
state, being an incompressible phase in the direct 
vicinity of the Laughlin phase.

\section*{Acknowledgements}
This work has been supported by EU (NAMEQUAM, AQUTE), 
ERC (QUAGATUA), Spanish MINCIN (FIS2008-00784 TOQATA), 
Alexander von Humboldt Stiftung, and AAII-Hubbard. 
B.~J.-D. is supported by the Ram\'on y Cajal program. 
M. L. acknowledges support from the Joachim Herz 
Foundation and Hamburg University.

\bibliographystyle{apsrev_mod}
\bibliography{bib.bib}
\end{document}